\documentclass[aps,showpacs,preprintnumbers,nofootinbib, epsf]{revtex4}
\usepackage{graphicx}

\begin{document}

\title{Viscous dissipative Chaplygin gas dominated homogenous and isotropic
cosmological models}
\author{C. S. J. Pun$^1$}
\email{jcspun@hkucc.hku.hk}
\author{L. \'{A}. Gergely$^{2,3,4}$}
\email{gergely@physx.u-szeged.hu}
\author{M. K. Mak$^1$}
\email{mkmak@vtc.edu.hk}
\author{Z. Kov\'{a}cs$^{3,5}$}
\email{zkovacs@titan.physx.u-szeged.hu}
\author{G. M. Szab\'{o}$^{3}$}
\email{szgy@titan.physx.u-szeged.hu}
\author{T. Harko$^1$}
\email{harko@hkucc.hku.hk}
\affiliation{$^{1}$Department of Physics and Center for Theoretical and Computational
Physics, The University of Hong Kong, Pok Fu Lam Road, Hong Kong}
\affiliation{$^{2}$Department of Theoretical Physics, University of Szeged, Tisza Lajos
krt 84-86, Szeged 6720, Hungary}
\affiliation{$^{3}$Department of Experimental Physics, University of Szeged, D\'{o}m T%
\'{e}r 9, Szeged 6720, Hungary}
\affiliation{$^{4}$Department of Applied Science, London South Bank University, 103
Borough Road, London SE1 OAA, UK}
\affiliation{$^{5}$Max-Planck-Institute f\"{u}r Radioastronomie, Auf dem H\"{u}gel 69,
53121 Bonn, Germany}
\date{\today }

\begin{abstract}
The generalized Chaplygin gas, which interpolates between a high
density relativistic era and a non-relativistic matter phase, is a
popular dark energy candidate. We consider a generalization of the
Chaplygin gas model, by assuming the presence of a bulk viscous
type dissipative term in the effective thermodynamic pressure of
the gas. The dissipative effects are described by using the
truncated Israel-Stewart model, with the bulk viscosity
coefficient and the relaxation time functions of the energy
density only. The corresponding cosmological dynamics of the bulk
viscous Chaplygin gas dominated universe is considered in detail
for a flat homogeneous isotropic Friedmann-Robertson-Walker
geometry. For different values of the model parameters we consider
the evolution of the cosmological parameters (scale factor, energy
density, Hubble function, deceleration parameter and luminosity
distance, respectively), by using both analytical and numerical
methods. In the large time limit the model describes an
accelerating universe, with the effective negative pressure
induced by the Chaplygin gas and the bulk viscous pressure driving
the acceleration. The theoretical predictions of the luminosity
distance of our model are compared with the observations of the
type Ia supernovae. The model fits well the recent supernova data.
From the fitting we determine both the equation of state of the
Chaplygin gas, and the parameters characterizing the bulk
viscosity. The evolution of the scalar field associated to the
viscous Chaplygin fluid is also considered, and the corresponding
potential is obtained. Hence the viscous Chaplygin gas model
offers an effective dynamical possibility for replacing the
cosmological constant, and to explain the recent acceleration of
the universe.
\end{abstract}

\maketitle



\section{Introduction}

The observations of high redshift supernovae \cite{Pe99} and the
Boomerang/Maxima/WMAP data \cite{Ber00}, showing that the location
of the first acoustic peak in the power spectrum of the microwave
background radiation is consistent with the inflationary
prediction $\Omega =1$, have provided compelling evidence for a
net equation of state of the cosmic fluid lying in the range
$-1\leq w=p/\rho <-1/3$. To explain these observations, two dark
components are invoked: the pressureless cold dark matter (CDM)
and the dark energy (DE) with negative pressure. CDM contributes
$\Omega _{m}\sim 0.3$, and is mainly motivated by the theoretical
interpretation of the galactic rotation curves and large scale
structure formation. DE is assumed to provide $\Omega _{DE}\sim
0.7$ and is responsible for the acceleration of the distant type
Ia supernovae. There are a huge number of candidates for DE in the
literature (for recent reviews see \cite{PeRa03} and \cite{Pa03}).

One possibility are cosmologies based on a mixture of cold dark
matter and quintessence, a slowly-varying, spatially inhomogeneous
component \cite{8}. An example of implementation of the idea of
quintessence is the suggestion that it is the energy associated
with a scalar field $Q$ with self-interaction potential $V(Q)$. If
the potential energy density is greater than the kinetic one, then
the pressure $p=\dot{Q}^{2}/2-V(Q)$ associated to the $Q$-field is
negative. Quintessential cosmological models have been intensively
investigated in the physical literature \cite{quint}.

A different line of thought has been followed in \cite {15,16,17},
where the conditions under which the dynamics of a self-interacting Brans$%
-$Dicke (BD) field can account for the accelerated expansion of
the Universe have been analyzed. Accelerated expanding solutions
can be obtained with a quadratic self-coupling of the BD field and
a negative coupling constant $\omega $ \cite{15}.

Dissipative effects, including both bulk and shear viscosity, are
supposed to play a very important role in the early evolution of
the Universe. A cosmic fluid (pressureless and with pressure)
obeying a perfect fluid type equation of state cannot support the
acceleration \cite{17}. A solution to this problem, and thus
avoiding the necessity of a potential for the BD field, is to
assume that some dissipative effects of bulk viscous type take
place at the cosmological scale \cite{16}. A combination of a
cosmic fluid with bulk dissipative pressure and quintessence
matter can drive an accelerated expansion phase of the Universe
and also solve the coincidence problem (the observational fact
that the energy density of cold dark matter and of $Q$-matter
should be comparable today) \cite{11}. The dynamics of a causal
bulk viscous cosmological fluid filled flat homogeneous Universe
in the framework of the BD theory was considered in \cite{MaHa03}.
The bulk viscous pressure term in the matter energy-momentum
tensor leads to a non-decelerating evolution of the Universe.

Neither CDM nor DE have direct laboratory observational or
experimental evidence for their existence. Therefore it would be
important if a unified dark matter - dark energy scenario could be
found, in which these two components are different manifestations
of a single fluid \cite{Paetal}. A candidate for such an
unification is the so-called generalized Chaplygin gas, which is
an exotic fluid with the equation of state $p=-B/\rho ^{n}$, where
$B$ and $n$ are two parameters to be determined. It was initially
suggested in \cite{Ka01} with $n=1$, and then generalized in
\cite{Be02} for the case $n\neq 1$. The Chaplygin gas also appears
in the stabilization of branes in
Schwarzschild-AdS black hole bulks as a critical theory at the horizon \cite%
{Ka00} and in the stringy analysis of black holes in three dimensions \cite%
{Ka98}. The Chaplygin equation of state can be derived from Born-Infeld type
Lagrangians \cite{Be02}, \cite{No05}. This simple and elegant model smoothly
interpolates between a non-relativistic matter phase ($p=0$) and a
negative-pressure dark energy dominated phase.

The cosmological implications of the Chaplygin gas model have been
intensively investigated in the recent literature \cite{Chco}. The Chaplygin
gas cosmological model has been constrained by using different cosmological
observations, like type Ia supernovae \cite{Fa02}, the CMB anisotropy
measurements \cite{Be03}, gravitational lensing surveys \cite{De03}, the age
measurement of high redshift objects \cite{Al03} and the X-ray gas mass
fraction of clusters \cite{Cu04}. The obtained results are somewhat
controversial, with some of them claiming good agreement between the data
and the Chaplygin gas model, while the rest ruling it as a feasible
candidate for dark matter. In particular, the standard Chaplygin gas model
with $n=1$ is ruled out by the data at a 99\% level \cite{Cu04}. The exact
solutions of the gravitational field equations in the generalized
Randall-Sundrum model for an anisotropic brane with Bianchi type I geometry,
with a generalized Chaplygin gas as matter source were obtained in \cite%
{MaHa05}.

The possibility of constraining Chaplygin dark energy models with
current Integrated Sachs Wolfe (ISW) effect data was investigated
in \cite{GiMe06}. In the case of a flat universe the generalized
Chaplygin gas models must have an energy density such that $\Omega
_{c}>0.55$ and an equation of state $w<-0.6$ at 95\% confidence
level. The extent to which the knowledge of spatial topology may
place constraints on the parameters of the generalized Chaplygin
gas (GCG) model for unification of dark energy and dark matter was
studied in \cite{Ber06}. By using both the Poincar\'{e}
dodecahedral and binary octahedral spaces as the observable
spatial topologies, the current type Ia supernovae (SNe Ia)
constraints on the GCG model parameters were
examined. An action formulation for the GCG model was developed in \cite%
{BaGhKu07}, and the most general form for the nonrelativistic GCG
action consistent with the equation of state has been derived. The
thermodynamical properties of dark energy have been investigated
in \cite{GoWaWa07}. For dark energy with constant equation of
state $w>-1$ and the generalized Chaplygin gas, the entropy is
positive and satisfies the entropy bound. Observational
constraints on the generalized Chaplygin gas (GCG) model for dark
energy from the 9 Hubble parameter data points, the 115 SNLS Sne
Ia data and the size of baryonic acoustic oscillation peak at
redshift, $z=0.35$ were examined in \cite{WuYu07}. At a
95.4\% confidence level, a combination of the three data sets gives $%
0.67\leq B/\rho_0^{1+n}\leq 0.83$ (where $\rho_0$ is the present day energy
density) and $-0.21\leq n\leq 0.42$, which is within the allowed
parameters ranges of the GCG as a candidate of the unified dark matter and
dark energy. However, the standard Chaplygin gas model ($n=1$) is also ruled
out by these data at the 99.7\% confidence level. A geometrical explanation
for the generalized Chaplygin gas within the context of brane world
theories, where matter fields are confined to the brane by means of the
action of a confining potential, was considered in \cite{HeSe07}.

The evolution of the Universe contains a sequence of important dissipative
processes, including GUT (Grand Unified theory) phase transition, taking
place at $t\approx 10^{-34}$ s and a temperature of about $T\approx 10^{27}$
K, when gauge bosons acquire mass, reheating of the Universe at the end of
inflation ($t\approx 10^{-32}$ s), when the scalar field decays into
particles, decoupling of neutrinos from the cosmic plasma ($t\approx 1$ s, $%
T\approx 10^{10}$ K), when the temperature falls below the threshold for
interactions that keep the neutrinos in thermal contact, nucleosynthesis,
decoupling of photons from matter during the recombination era ($t\approx 10$
s, $T\approx 10^{3}$ K), when electrons combine with protons and no longer
scatter the photons etc. \cite{Ma}.

The first attempts at creating a theory of relativistic
dissipative fluids were those of Eckart \cite{Ec40} and Landau and
Lifshitz \cite{LaLi87}. These theories are now known to be
pathological in several respects. Regardless of the choice of the
equation of state, all equilibrium states in these theories are
unstable and in addition signals may be propagated through the
fluid at velocities exceeding the speed of light. These problems
arise due to the first order nature of the theory, that is, it
considers only first-order deviations from the equilibrium leading
to parabolic differential equations, hence to infinite speeds of
propagation for heat flow and viscosity, in contradiction with the
principle of causality. Conventional theory is thus applicable
only to phenomena which are quasi-stationary, i.e. slowly varying
on space and time scales characterized by mean free path and mean
collision time.

A relativistic second-order theory was found by Israel \cite{Is76} and
developed in \cite{IsSt76} and  \cite%
{HiLi89,HiSa91} into what is called ``transient'' or ``extended''
irreversible thermodynamics. In this model deviations from
equilibrium (bulk stress, heat flow and shear stress) are treated
as independent dynamical variables, leading to a total of 14
dynamical fluid variables to be determined. For general reviews on
causal thermodynamics and its role in relativity see \cite{Ma} and
\cite{Ma95}. Causal bulk viscous thermodynamics has been
extensively used for describing the dynamics and evolution of the
early Universe, or in an astrophysical context \cite%
{ChJa97}.

It is the purpose of this paper to consider the effects of a
possible existence of a bulk viscosity of the generalized
Chaplygin gas on the cosmological dynamics of the Universe. The
viscous effects are described by using the truncated
Israel-Stewart theory \cite{IsSt76}. By using the Laplace
transformation and the convolution theorem, the second order
differential equation describing the evolution of the Hubble
parameter $H$ is transformed into an integral equation. The field
equations are solved by means of an iterative scheme. Then the
general solutions of the equations are obtained in a parametric
form in the zero, first, second and $m$th order approximation, and
the relevant cosmological parameters (scale factor, energy
density, Hubble parameter, deceleration parameter etc.) are
obtained. The scalar field interpretation of the Chaplygin gas is
generalized to take into account the viscosity and dissipative
effects.

In order to compare the predictions of the model with the
observational data we have fitted the luminosity distance-redshift
relation with the latest observational data of the type Ia
supernovae. The model fits well these data. From the fitting we
determine both the equation of state of the Chaplygin gas, and the
parameters characterizing the bulk viscosity. Even by taking into
account the effect of the bulk viscosity, the $n=1$ Chaplygin gas
models are ruled out by the observations.

The present paper is organized as follows. The physical model and
the basic equations are presented in Section II. The evolution
equation for the Hubble parameter is studied in Section III, and
the behavior of the cosmological parameters is obtained. The
observational data have been compared with the theoretical
predictions of the model in Section IV. In Section V we discuss
and conclude our results. In the present paper we use a system of
units so that $8\pi G=c=1$.

\section{Geometry, field equations and consequences}

Perfect fluids in equilibrium generate no entropy and no
frictional type heating, since their dynamics is reversible and
without dissipation. A perfect fluid model is adequate for the
description of many
processes in cosmology. However,\textit{\ real fluids} behave \textit{%
irreversibly}, and some processes in astrophysics and cosmology cannot be
understood except as irreversible processes. An important irreversible
effect is bulk viscosity, which typically arises in mixtures, either of
different species, as is the case of the radiative fluid, or of the same
species, but with different energies, as in a Maxwell-Boltzmann gas.
Physically, in cosmology we can think of bulk viscosity as an internal
friction due to the different cooling rates in an expanding gas. The
dissipation due to bulk viscosity converts kinetic energy of the particles
into heat, and thus we expect it to reduce the effective pressure in an
expanding fluid \cite{Ma,Ma95}.

For a flat homogeneous Friedmann-Robertson-Walker (FRW) with a line element:
\begin{equation}
ds^{2}=dt^{2}-a^{2}(t)\left( dx^{2}+dy^{2}+dz^{2}\right) ,
\end{equation}
filled with a bulk viscous cosmological fluid the energy-momentum tensor is
given by :
\begin{equation}
T_{i}^{k}=\left( \rho +p+\Pi \right) u_{i}u^{k}-\left( p+\Pi \right) \delta
_{i}^{k},  \label{1}
\end{equation}
where $\rho $ is the energy density, $p$ the thermodynamic pressure, $\Pi $
the bulk viscous pressure and $u_{i}$ the four velocity satisfying the
condition $u_{i}u^{i}=1$. The effect of the bulk viscosity of the
cosmological fluid can be considered by adding to the usual thermodynamic
pressure $p$ the bulk viscous pressure $\Pi $, and formally substituting the
pressure terms in the energy-momentum tensor by $p_{\text{eff}}=p+\Pi $. The
particle and entropy fluxes are defined according to $N^{i}=nu^{i}$ and $%
S^{i}=\sigma N^{i}-\left( \tau \Pi ^{2}/2\xi T\right) u^{i}$, with $n$ is
the number density, $\sigma $ is the specific entropy, $T\geq 0$ is the
temperature, $\xi $ is the bulk viscosity coefficient and $\tau \geq 0$ is
the relaxation coefficient for transient bulk viscous effect (i.e. the
relaxation time). The evolution of the cosmological fluid is subject to the
dynamical laws of particle number conservation $N_{\text{ };i}^{i}=0$ and
Gibb's equation $Td\sigma =d\left( \rho /n\right) +pd\left( 1/n\right) $
\cite{Ma,Ma95}. In the following we shall also suppose that the
energy-momentum tensor of the cosmological fluid is conserved, that is $%
T_{i;k}^{k}=0$, where $;$ denotes the covariant derivative with respect to
the metric..

The gravitational field equations together with the continuity equation $%
T_{i;k}^{k}=0$ imply
\begin{equation}
3H^{2}=\rho ,
\end{equation}
\begin{equation}
2\dot{H}+3H^{2}=-p-\Pi ,  \label{field}
\end{equation}
\begin{equation}
\dot{\rho}+3\left( \rho +p\right) H=-3H\Pi ,
\end{equation}
where $H=\dot{a}/a$ is the Hubble parameter.

For the evolution of the bulk viscous pressure $\Pi $ we adopt the truncated
evolution equation \cite{Ma,Ma95}, obtained in the simplest way (linear in $%
\Pi )$ to satisfy the $H$-theorem (i.e. for the entropy production to be
non-negative, $S_{;i}^{i}=\Pi ^{2}/\xi T\geq 0$ \cite{IsSt76}.The evolution
equation for $\Pi $ is given in the framework of the truncated
Israel-Stewart theory by \cite{Ma}
\begin{equation}
\tau \dot{\Pi}+\Pi =-3\xi H,  \label{bulk}
\end{equation}
where $\xi $ the bulk viscosity coefficient and $\tau $ the relaxation time.
The truncated equation is a good approximation of the full causal transport
equations if the condition $\left| \Pi d\left( a^{3}\tau /\xi T\right)
/dt\right| <<a^{3}H/T$ holds \cite{Ma,Ma95}. In order to close the system of
equations (\ref{field}) and (\ref{bulk}) we have to give the equation of
state for $p$ and specify $\tau $ and $\xi $.

We assume that the isotropic pressure $p$ of the cosmological fluid obeys a
modified Chapylin gas equation of state \cite{De04},
\begin{equation}  \label{chap}
p=\gamma \rho -\frac{B}{\rho ^{n}},
\end{equation}
where $0\leq \gamma \leq 1$ and $0\leq n \leq 1$. $B$ is a positive constant.

When $\gamma =1/3$ and the comoving volume of the Universe is small ($\rho
\rightarrow \infty $), this equation of state corresponds to a radiation
dominated era. When the density is small, $\rho \rightarrow 0$, the equation
of state corresponds to a cosmological fluid with negative pressure (the
dark energy). Generally the modified Chaplygin equation of state corresponds
to a mixture of ordinary matter and dark energy. For $\rho =\left( B/\gamma
\right) ^{1/(n+1)}$ the matter content is pure dust with $p=0$. The speed of
sound $v_{s}=\left( \partial p/\partial \rho \right) ^{1/2}$ in the
Chaplygin gas is given by
\begin{equation}
v_{s}^{2}=\gamma (1+n)-\frac{np}{\rho }.
\end{equation}

For the bulk viscosity coefficient and for the relaxation time of the
viscous Chaplygin gas we assume the following phenomenological laws
\begin{equation}
\xi =\alpha \rho ^{s},\qquad \tau =\xi \rho ^{-1}=\alpha \rho ^{s-1},
\label{csi}
\end{equation}%
where $0\leq \gamma \leq 1$, $\alpha \geq 0$ and $s\geq 0$ are constants
\cite{law}. Eqs. (\ref{csi}) are standard in cosmological models, whereas
the equation for $\tau $ is a simple procedure to ensure that the speed of
viscous pulses does not exceed the speed of light.

The truncated Israel-Stewart theory is derived under the
assumption that the thermodynamical state of the fluid is close to
equilibrium, that is the non-equilibrium bulk viscous pressure
should be small when compared to the local equilibrium pressure
$\left| \Pi \right| <<p=\gamma \rho -B/\rho ^{n}$. If this
condition is violated then one is effectively assuming that the
linear theory holds also in the nonlinear regime far from
equilibrium. However, for a fluid description of the matter, the
condition ought to be satisfied.

To see if a cosmological model accelerates or not it is convenient to
introduce the deceleration parameter
\begin{equation}
q=\frac{dH^{-1}}{dt}-1=\frac{\rho +3p+3\Pi }{2\rho
}=\frac{1}{2}+\frac{3\gamma
(n+1)}{2n}\left[1-\frac{v_s^2}{(n+1)\gamma }\right]+\frac{3\Pi
}{2\rho }.
\end{equation}

The positive sign of the deceleration parameter corresponds to standard
decelerating models whereas the negative sign indicates accelerated
expansion.

By using the assumptions given by Eqs. (\ref{csi}) for the bulk viscosity
coefficient and the relaxation time, the evolution equation for the Hubble
parameter $H$ for the viscous dissipative Chaplygin gas dominated flat
homogeneous cosmological models is obtained from the field equations as
\begin{equation}
\ddot{H}+\left[ 3\left( \gamma +1\right) H+\frac{nB}{3^{n}}H^{-2n-1}+\frac{%
3^{1-s}}{\alpha }H^{2-2s}\right] \dot{H}-\frac{3^{1-n-s}}{2\alpha }%
BH^{2-2s-2n}+\frac{3^{2-s}\left( \gamma +1\right) }{2\alpha }H^{4-2s}-\frac{9%
}{2}H^{3}=0.  \label{ev}
\end{equation}

\section{Iterative solutions of the evolution equation}

In order to obtain a simpler form of Eq. (\ref{ev}) we introduce
the dimensionless functions $\theta $ and $h$ by means of the
definitions
\begin{equation}
H=\left( 3^{s}\alpha \right) ^{\frac{1}{1-2s}}h,\theta =\frac{3}{\sqrt{2}}%
\left( 3^{s}\alpha \right) ^{\frac{1}{1-2s}}t,s\neq \frac{1}{2},
\label{transfor}
\end{equation}
and we denote $\lambda _{0}=(B/3^{n+1})\left( 3^{s}\alpha \right)
^{-2\left( 1+n\right)/(1-2s)}$, $s\neq 1/2$.

In these variables Eq. (\ref{ev}) takes the form
\begin{equation}
\frac{d^{2}h}{d\theta ^{2}}+\sqrt{2}\left[ \left( \gamma +1\right)
h+n\lambda _{0}h^{-1-2n}+h^{2\left( 1-s\right) }\right] \frac{dh}{d\theta }%
-\lambda _{0}h^{2\left( 1-s-n\right) }+\left( \gamma +1\right)
h^{2\left( 2-s\right) }-h^{3}=0,s\neq \frac{1}{2}.  \label{ev1}
\end{equation}

Introducing the new variables $h=\sqrt{y}$ and $\eta =\int
\sqrt{y}d\theta $, respectively, Eq. (\ref{ev1}) becomes
\begin{equation}
\frac{d^{2}y}{d\eta ^{2}}+\sqrt{2}\left( \gamma +1\right) \frac{dy}{d\eta }%
-2y+\sqrt{2}\left( n\lambda _{0}y^{-n-1}+y^{\frac{1}{2}-s}\right) \frac{dy}{%
d\eta }-2\lambda _{0}y^{\frac{1}{2}-s-n}+2\left( \gamma +1\right) y^{\frac{3%
}{2}-s}=0,s\neq \frac{1}{2}.  \label{final}
\end{equation}

Therefore for a fixed equation of state and known values of $s$
and $n$ the evolution of the viscous Chaplygin gas cosmological
models is determined by a single numerical parameter $\lambda
_{0}$.

Due to the complicated non-linear character of the evolution
equation (\ref {final}), it is very difficult to obtain exact
solutions of this equation in the framework of the truncated
Israel-Stewart theory. The cosmological model
presented above, could be robust if the cosmological solutions of equation (%
\ref{final}), depicting the causal bulk viscous FRW space-time,
could be studied for an arbitrary range of values of $s$, $n$ and
$\gamma $ in the hope of leading to the possibility of correct
physical description of a well-determined period in the evolution
of our Universe. By using the Laplace transform and convolution
theorem, the differential equation (\ref {final}) is equivalent
with the following integral equation
\begin{equation}
y\left( \eta \right) =\int_{0}^{\eta }F\left( \eta -x\right) \left[ \sqrt{2}%
\left( n\lambda _{0}y^{-n-1}+y^{\frac{1}{2}-s}\right) \frac{dy}{d\eta }%
-2\lambda _{0}y^{\frac{1}{2}-s-n}+2\left( \gamma +1\right) y^{\frac{3}{2}-s}%
\right] dx+y_{0}\left( \eta \right) ,  \label{inteq}
\end{equation}
where
\begin{equation}
F\left( \eta -x\right) =\frac{1}{2\delta }\left[ e^{-\left( \delta +\frac{%
\left( \gamma +1\right) }{\sqrt{2}}\right) \left( \eta -x\right)
}-e^{\left( \delta -\frac{\left( \gamma +1\right)
}{\sqrt{2}}\right) \left( \eta -x\right) }\right] ,
\end{equation}
\begin{equation}
y_{0}\left( \eta \right) =e^{-\frac{\left( \gamma +1\right)
}{\sqrt{2}}\eta }\left( Me^{-\delta \eta }+Ne^{\delta \eta
}\right) ,
\end{equation}
\begin{equation}
M=\frac{n_{+}y(0)-y^{\ast }(0)}{2\delta },N=\frac{y^{\ast }(0)-n_{-}y(0)}{%
2\delta },
\end{equation}
and we denoted $\delta =\sqrt{\gamma ^{2}+2\gamma +5}/\sqrt{2}$, $n_{\pm }=-\left( \gamma +1\right) /\sqrt{2}\pm \delta $ and $%
y^{\ast }(0)=\left( dy/d\eta \right) _{\eta =0}$, respectively.

The solution of the integral equation (\ref{inteq}) can be easily
obtained by using the method of successive approximations or
method of iteration to obtain a solution to any desired accuracy.
Taking as an initial approximation the solution of the linear part
of equation (\ref{final}), the general solution of the integral
equation (\ref{inteq}) can be expressed in the first and $m$th
order approximation, $m\in N$, as follows
\begin{equation}
y_{1}\left( \eta \right) =\int_{0}^{\eta }F\left( \eta -x\right)
\left[
\sqrt{2}\left( n\lambda _{0}y_{0}^{-n-1}\left( x\right) +y_{0}^{\frac{1}{2}%
-s}\left( x\right) \right) y_{0}^{^{\prime }}\left( x\right)
-2\lambda
_{0}y_{0}^{\frac{1}{2}-s-n}\left( x\right) +2\left( \gamma +1\right) y_{0}^{%
\frac{3}{2}-s}\left( x\right) \right] dx+y_{0}\left( \eta \right)
,
\end{equation}
\begin{eqnarray*}
&&.\text{ \ }.\text{ \ }. \\
&&.\text{ \ }.\text{ \ }.
\end{eqnarray*}
\begin{equation}
y_{m}\left( \eta \right) =\int_{0}^{\eta }F\left( \eta -x\right)
\left[
\sqrt{2}\left( n\lambda _{0}y_{m-1}^{-n-1}\left( x\right) +y_{m-1}^{\frac{1}{%
2}-s}\left( x\right) \right) y_{m-1}^{^{\prime }}\left( x\right)
-2\lambda _{0}y_{m-1}^{\frac{1}{2}-s-n}\left( x\right) +2\left(
\gamma +1\right) y_{m-1}^{\frac{3}{2}-s}\left( x\right) \right]
dx+y_{m-1}\left( \eta \right) ,
\end{equation}
\begin{equation}
y\left( \eta \right) =\lim_{m\rightarrow \infty }y_{m}\left( \eta
\right) .
\end{equation}

We can express the iterative solutions of the gravitational field
equations for a bulk-viscous fluid filled FRW Universe in the
framework of the truncated Israel-Stewart theory for $s\neq 1/2$
in the following parametric form (in the following equations we
write $\sigma $ for the variable of integration in order to
distinguish it from the independent variable):
\begin{equation}
\theta -\theta _{0}=\int_{\eta _{0}}^{\eta }\frac{1}{\sqrt{y\left(
\sigma \right) }}d\sigma ,a=a_{0}e^{\frac{\sqrt{2}}{3}\eta },
\end{equation}
\begin{equation}
\rho =\rho _0 y\left( \eta \right) , p=\rho _0\left[ \gamma
y\left( \eta \right) -\frac{\lambda _0}{y^{n}\left( \eta \right)
}\right] ,
\end{equation}
\begin{equation}
\xi =\sqrt{\frac{\rho _0}{3}}y^{s}\left( \eta \right) ,\tau
=\sqrt{\frac{\rho _0}{27}}y^{s-1}\left( \eta \right) ,
\end{equation}
\begin{equation}
q=-\frac{3}{2\sqrt{2}}\frac{1}{y\left( \eta \right)
}\frac{dy}{d\eta }-1,
\end{equation}
\begin{equation}
\Pi =\rho _0\left[ -\left( \gamma
+1\right) y(\eta )+\frac{\lambda _{0}}{y^{n}\left( \eta \right) }-\frac{1}{%
\sqrt{2}}\frac{dy}{d\eta }\right] ,
\end{equation}
where $a_{0}$, and $t_{0}$ are arbitrary constants of integration,
and we denoted $\rho _0 =3\left( 3^{s}\alpha \right) ^{2/(1-2s)}$.

In order to solve the evolution equation iteratively we need to
chose the initial conditions for the cosmological model. The
initial value of the function $y\left( \eta \right) $ can be
obtained by fixing the initial value of the Hubble function or,
equivalently, of the density, by using the
equation $y\left( 0\right) =3%
H^{2}\left( 0\right)/\rho _0 =\rho \left( 0\right) /\rho _0$. The
initial value of $\left. dy/d\eta \right| _{\eta =0}$ can be
obtained by fixing the initial value of the deceleration
parameter, so that $\left. dy/d\eta \right| _{\eta =0}=-\left(
2\sqrt{2/3}\right) \left[ q\left( 0\right) +1\right] $. In this
way the mathematical initial conditions are fixed by the physical
characteristics of the Universe.

The behavior of the cosmological parameters of the bulk viscous
Chaplygin gas filled homogeneous and isotropic universe dust universe, with $\gamma =0$%
, and, consequently, $p=-B/\rho ^{n}$, are represented, for some
fixed values of $n$ and $s$ and for different values of $\lambda
_{0}$ in Figs.~\ref{fig1}-\ref{fig6}. In Fig.~{\ref{fig1} the
evolution of the scale factor $a$ is represented as a function of
the dimensionless time $\theta $.

\begin{figure}[th]
\centering
\includegraphics{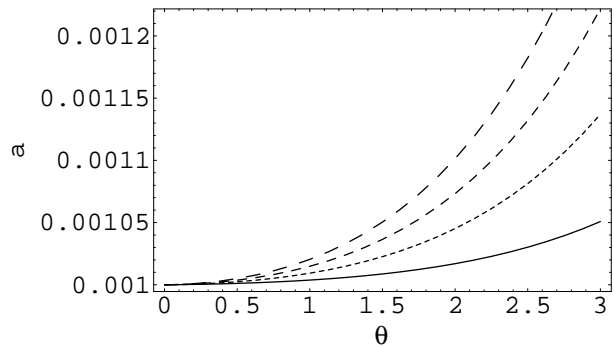}
\caption{The scale factor $a$ of the bulk viscous Chaplygin gas
filled homogeneous and isotropic dust universe ($\gamma =0$) as a
function of the dimensionless time $\theta =\alpha ^{1/(1-2s)}t$,
for
$n=0.1$, $s=1/4$ and different values of $\lambda _0$: $%
\lambda _0=0.01$ (solid curve), $\lambda _0=0.03$ (dotted curve), $%
\lambda _0=0.05$ (dashed curve) and $\lambda _0=0.07$ (long dashed
curve).} \label{fig1}
\end{figure}


The time variation of the density of the matter is plotted against
the time in Fig.~\ref{fig3}. In the expanding universe the density
is a monotonically decreasing function of the cosmic time.

\begin{figure}[!ht]
\centering
\includegraphics{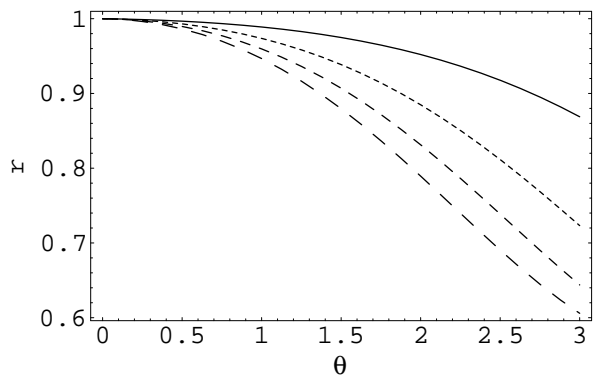}
\caption{The dimensionless density function $r=\rho /\alpha
^{2/(1-2s)}$ of the bulk viscous Chaplygin gas filled homogeneous
and isotropic dust universe ($\gamma =0$) as a function of the
dimensionless time $\theta =\protect\alpha ^{1/(1-2s)}t$, for $n=0.1$%
, $s=1/4$ and different values of $\lambda _0$: $%
\lambda _0=0.01$ (solid curve), $\lambda _0=0.03$ (dotted curve), $%
\lambda _0=0.05$ (dashed curve) and $\lambda _0=0.07$ (long dashed
curve).} \label{fig3}
\end{figure}

The behavior of the bulk viscous pressure $\Pi $ of the Chaplygin
gas is shown in Fig.~\ref{fig4}. The negative bulk viscous
pressure gives a significant contribution to the total negative
pressure of the Chaplygin gas.

\begin{figure}[!ht]
\centering
\includegraphics{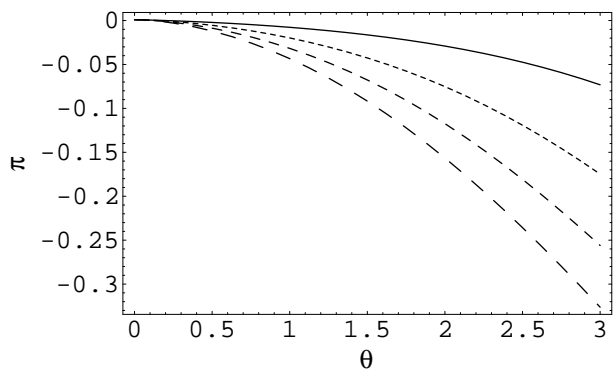}
\caption{The dimensionless bulk viscous pressure $\pi=\Pi /\protect%
\alpha ^{2/(1-2s)}$ of the Chaplygin gas for a homogeneous and
isotropic
dust universe ($\gamma =0$) as a function of the dimensionless time $%
\theta =\alpha ^{1/(1-2s)}t$, for $n=0.1$, $s=1/4$ and different
values of $\lambda _0$: $%
\lambda _0=0.01$ (solid curve), $\lambda _0=0.03$ (dotted curve), $%
\lambda _0=0.05$ (dashed curve) and $\lambda _0=0.07$ (long dashed
curve).} \label{fig4}
\end{figure}

The time variation of the bulk viscosity coefficient of the
Chaplygin gas is represented in Fig.~\ref{fig5}. Similarly to the
energy density, the bulk viscosity coefficient is a monotonically
decreasing function of time.

\begin{figure}[!ht]
\centering
\includegraphics{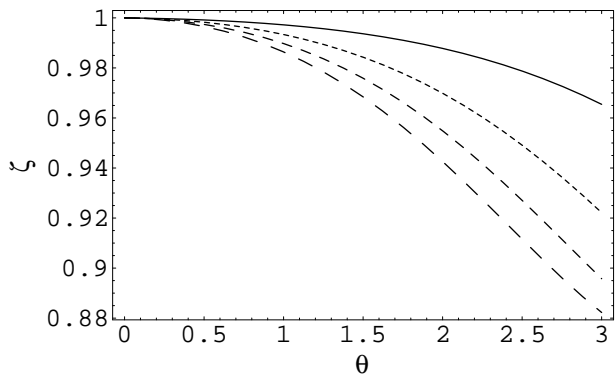}
\caption{The dimensionless bulk viscosity coefficient $\zeta=%
\xi /\alpha $ of the Chaplygin gas for a homogeneous and isotropic
dust universe ($\gamma =0$) as a function of the dimensionless time $%
\theta =\alpha ^{1/(1-2s)}t$, for $n=0.1$, $s=1/4$ and different
values of $\lambda _0$: $%
\lambda _0=0.01$ (solid curve), $\lambda _0=0.03$ (dotted curve), $%
\lambda _0=0.05$ (dashed curve) and $\lambda _0=0.07$ (long dashed
curve).} \label{fig5}
\end{figure}

The time variation of the deceleration parameter $q$ is
represented in Fig.~\ref{fig6}. In the limit of the large times
$q<0$, showing that the viscous Chaplygin gas filled universe
experiences an accelerated cosmological dynamics. For large values
of $\theta $, after experiencing a super-accelerated phase with
$q<-1$, the viscous Chaplygin gas filled universe ends in a de
Sitter regime, with $q=-1$.

\begin{figure}[!ht]
\centering
\includegraphics{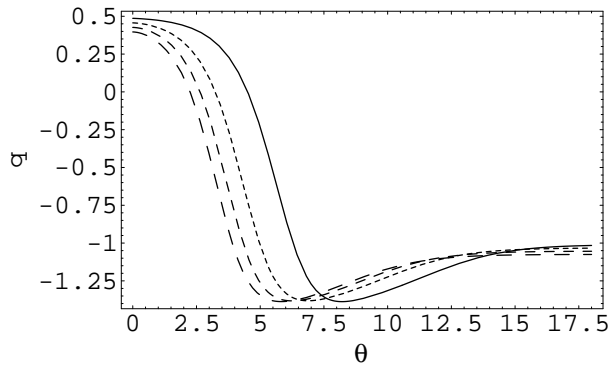}
\caption{The deceleration parameter $q$ of the bulk viscous
Chaplygin gas filled homogeneous and isotropic dust universe
($\gamma =0$) as a function of the dimensionless time $\theta
=\alpha ^{1/(1-2s)}t$, for $n=0.1$, $s=1/4$ and different values
of $\lambda _0$: $%
\lambda _0=0.01$ (solid curve), $\lambda _0=0.03$ (dotted curve), $%
\lambda _0=0.05$ (dashed curve) and $\lambda _0=0.07$ (long dashed
curve).} \label{fig6}
\end{figure}

When the bulk viscosity coefficient $\xi $ is proportional to the
square root of the density, $\xi \sim \rho ^{1/2}$, that is, for
$s=1/2$, the transformations introduced in Eqs. (\ref{transfor})
cannot be applied. In this case a set of dimensionless variable is
given by
\begin{equation}
H=\left( \frac{3^{n+1}}{B}\right) ^{-\frac{1}{2(1+n)}}h,\theta =\frac{3}{%
\sqrt{2}}\left( \frac{3^{n+1}}{B}\right) ^{-\frac{1}{2(1+n)}}t,
\end{equation}
while the dynamics of the Universe is determined by the parameter $\chi =1/%
\sqrt{3}\alpha $. In these variables and for $s=1/2$ the evolution equation (%
\ref{ev}) takes the form
\begin{equation}
\frac{d^{2}h}{d\theta ^{2}}+\sqrt{2}\left[ \left( \gamma +1+\chi
\right) h+nh^{-1-2n}\right] \frac{dh}{d\theta }-\chi
h^{1-2n}+\left[ \left( \gamma +1\right) \chi -1\right]
h^{3}=0,s=\frac{1}{2}.  \label{s1}
\end{equation}

The transformations $h=\sqrt{y}$ and $\eta =\int \sqrt{y}d\theta $
reduces Eq. (\ref{s1}) to
\begin{equation}
\frac{d^{2}y}{d\eta ^{2}}+\sqrt{2}\left[ \gamma +1+\chi
+ny^{-1-n}\right] \frac{dy}{d\eta }-2\chi y^{-n}+2\left[ \left(
\gamma +1\right) \chi -1\right] y=0.  \label{s2}
\end{equation}

The general behavior of the viscous Chaplygin gas models with
$s=1/2$ is qualitatively similar to the case $s\neq 1/2$.
Therefore we present only the time evolution of the deceleration
parameter $q$, which is shown in Fig.~\ref{fig7}.

\begin{figure}[!ht]
\centering
\includegraphics{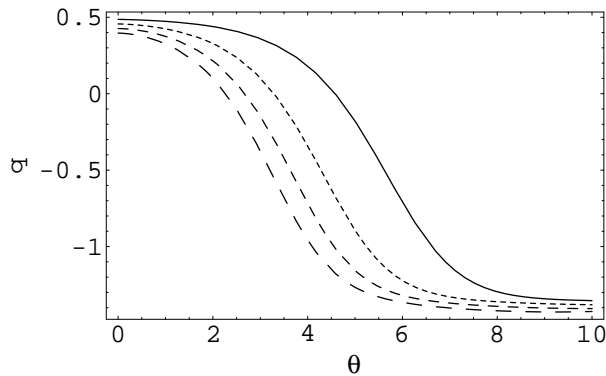}
\caption{The deceleration parameter $q$ of the homogeneous and
isotropic dust universe ($\protect\gamma =0$) filled with a
viscous Chaplygin gas,
with bulk viscosity proportional to the square root of the energy density ($%
s=1/2$) as a function of the dimensionless time $\protect\theta =t/sqrt{%
\protect\alpha }$, for $n=0.1$ and different values of $\lambda _0$: $%
\lambda _0=0.01$ (solid curve), $\lambda _0=0.03$ (dotted curve), $%
\lambda _0=0.05$ (dashed curve) and $\lambda _0=0.07$ (long dashed
curve).} \label{fig7}
\end{figure}

In the limit of large times the viscous Chaplygin universe with
$s=1/2$ ends in a super-accelerated state, with $q\approx-1.25$.

\section{Comparison with observational data}

From observational point of view fundamental tests of cosmological
models can be performed from the study of the propagation in a
curved space-time of the light emitted by a source in a distant
galaxy (like, for example, a supernova), and detected on Earth, on
a telescope mirror. The luminosity of the source is defined as
$L=dE_{em}/dt_{em}$, that is, the luminosity is the total energy
emitted by the source in unit time; the suffix $em$ refers to
emission. A telescope detects a photon flux
$F=dE_{rec}/dt_{rec}/A_{M}$, where the suffix $rec$ refers to
reception. The flux is the energy detected on the telescope mirror
surface $A_{M}$ (assumed to be perpendicular to the incident light
beam) per unit time interval \cite{ger}.

An important observational parameter, the redshift $z$ is defined as $%
1+z=a_{0}/a$, where $a_{0}$ is the present day value of the scale
factor, which is usually conventionally taken as $1$, $a_{0}=1$.
From the definition of $z$ we obtain $da/dt=-\left[ a_{0}/\left(
1+z\right) ^{2}\right] dz/dt$. From the definition of the Hubble
function we have $H=(1/a)(da/dt)=-[1/(1+z)](dz/dt)$, which gives
$dz/dt=-\left( 1+z\right) H$. Due to the cosmological expansion
the elementary area changes as $a^{2}$ and the frequency $\omega $
of the light is redshifted during the cosmic evolution so that
$\omega \propto 1/a$. Therefore $F/L=\left( 1/A_{tot}\right)
\left( a/a_{0}\right) ^{2}$, where $A_{tot}$ represents the proper
area of a sphere centered in the light source and containing at
the time of reception the reception point on its surface. The
luminosity distance is defined as \cite{ger}
\begin{equation}
d_{L}\left( z\right) =\sqrt{\frac{L}{4\pi F}}=a_{0}r_{em}\left(
1+z\right) ,
\end{equation}
where $r_{em}$ is the comoving radius. The comoving coordinate
$r_{em}$ can be written in terms of an other radial comoving
coordinate $\chi _{em}$, so that
\begin{equation}
\chi _{em}=\chi \left( a_{em}\right) =\int_{a_{em}}^{a_{0}}\frac{da}{%
a^{2}H(a)}=\frac{1}{a_{0}}\int_{0}^{z}\frac{dz^{\prime }}{H\left(
z^{\prime }\right) }.
\end{equation}
In a flat ($k=0$) FRW geometry we have $%
d_{L}(z)=a_{0}\left( 1+z\right) \chi _{em}$. The luminosity
distance-redshift relation is given by \cite{ger}
\begin{equation}
d_{L}\left( z\right) =\left( 1+z\right) \int_{0}^{z}\frac{dz^{\prime }}{%
H\left( z^{\prime }\right) }.
\end{equation}

In the case of a bulk viscous Chaplygin gas filled universe the
luminosity distance $d_{L}(z)$ can be obtained by simultaneously
solving the following system of differential equations, with $z$
as independent variable
\begin{equation}
-2(1+z)H\frac{dH}{dz}+3H^{2}=-\gamma \rho +\frac{B}{\rho ^{n}}-\Pi
, \label{eq1}
\end{equation}
\begin{equation}
-\left( 1+z\right) \frac{d\rho }{dz}+3\left[ \left( 1+\gamma \right) \rho -%
\frac{B}{\rho ^{n}}\right] =-3\Pi ,  \label{eq2}
\end{equation}
\begin{equation}
-\alpha \left( 1+z\right) \rho ^{s-1}H\frac{d\Pi }{dz}+\Pi
=-3\alpha \rho ^{s}H,  \label{eq3}
\end{equation}
and
\begin{equation}
\frac{dd_{L}(z)}{dz}-\frac{1}{1+z}d_{L}(z)=\frac{1+z}{H(z)},
\label{eq4}
\end{equation}
respectively. In order to simplify this system we introduce a set
of dimensionless variables, defined as
\begin{equation}
H(z)=H_{0}h(z),\rho (z)=3H_{0}^{2}r(z),\Pi (z)=3H_{0}^{2}\pi (z),d_{L}(z)=%
\frac{D_{L}(z)}{H_{0}},
\end{equation}
we denote $\lambda =B/3^{n}H_{0}^{2n+2}$, and choose $\alpha $ so
that $3^{s-1}\alpha H_{0}^{2s-1}=1$. Substitution into Eqs.
(\ref{eq1})-(\ref{eq4}) transform these equations into the form
\begin{equation}
-2(1+z)h\frac{dh}{dz}+3h^{2}=-3\gamma r+\frac{\lambda
}{r^{n}}-3\pi , \label{z1}
\end{equation}
\begin{equation}
-\left( 1+z\right) \frac{dr}{dz}+3\left[ \left( 1+\gamma \right) r-\frac{%
\lambda }{3r^{n}}\right] =-3\pi ,  \label{z2}
\end{equation}
\begin{equation}
-\left( 1+z\right) r^{s-1}h\frac{d\pi }{dz}+\pi =-3r^{s}h,
\label{z3}
\end{equation}
and
\begin{equation}
\frac{dD_{L}(z)}{dz}-\frac{1}{1+z}D_{L}(z)=\frac{1+z}{h(z)},
\label{z4}
\end{equation}
respectively. The initial conditions for the system of Eqs.
(\ref{z1})-(\ref {z4}) are $h(0)=1$, $r(0)=1$, $\pi (0)=\pi _{0}$
and $D_{L}(0)=0$, respectively. In the equation of state one can
take $\gamma =0$. The physical luminosity distance can be written
as $d_{L}(z;\lambda ;n;s)=H_{0}^{-1}D_{L}(z;\lambda ;n;s)$. Once
the dimensionless function $D_{L}(z;\lambda ;n;s)$ is known from
the numerical integration of the system, the fitting
with the observational data will fix the numerical values of the parameters $%
\lambda ,n,s$.

The function $d_{L}\left( z\right) $ can be measured for distant
type Ia supernovae. The luminosity is evaluated by photometry,
while the redshift is evaluated from the spectroscopic analysis of
the host galaxy. Each cosmological model has its own prediction
for the function $d_{L}\left( z\right) $. Therefore the measured
$d_{L}\left( z\right) $ data are powerful tests of the
cosmological models, and the luminosity distance can be used to
fit the free parameters of the model by using the observational
results.

\begin{figure}[tbp]
\includegraphics[width=5.5cm, angle=270]{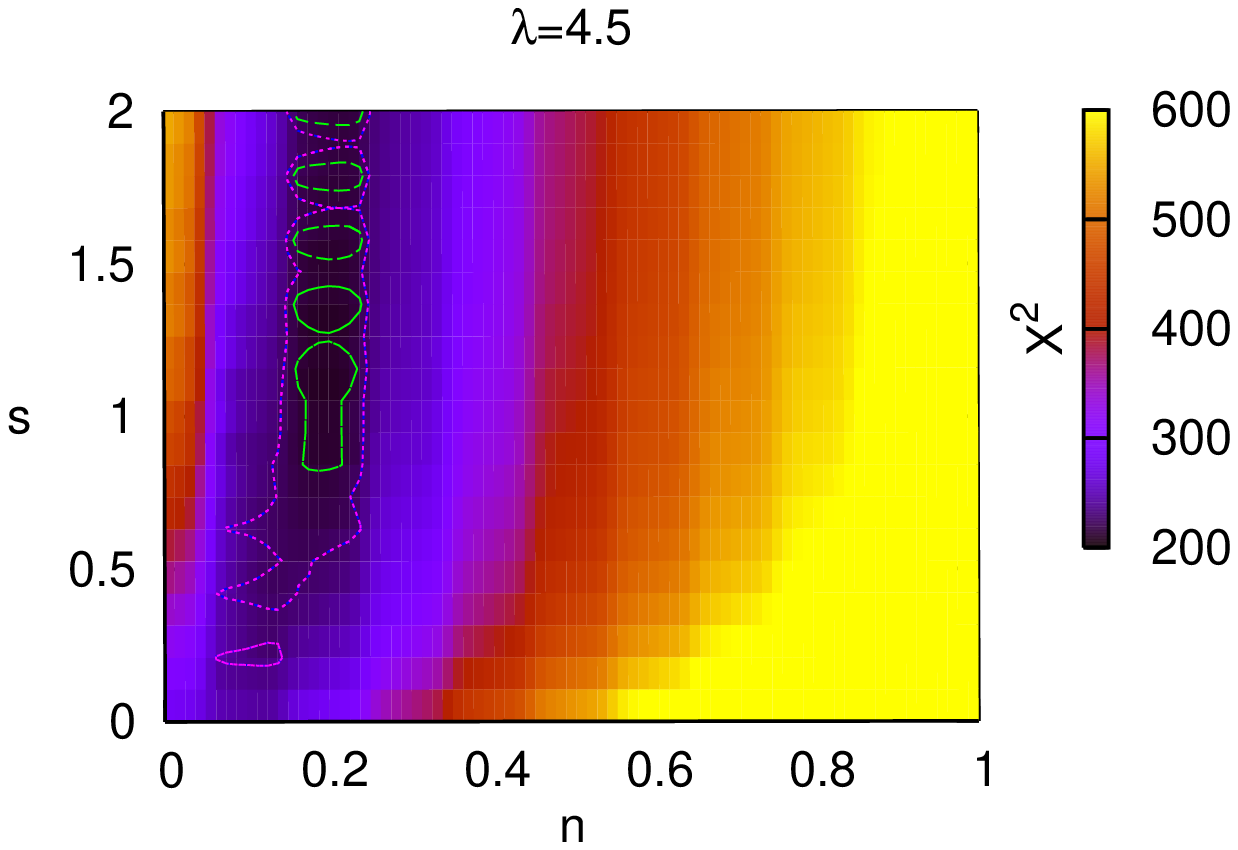}\hskip1cm%
\includegraphics[width=5.5cm, angle=270]{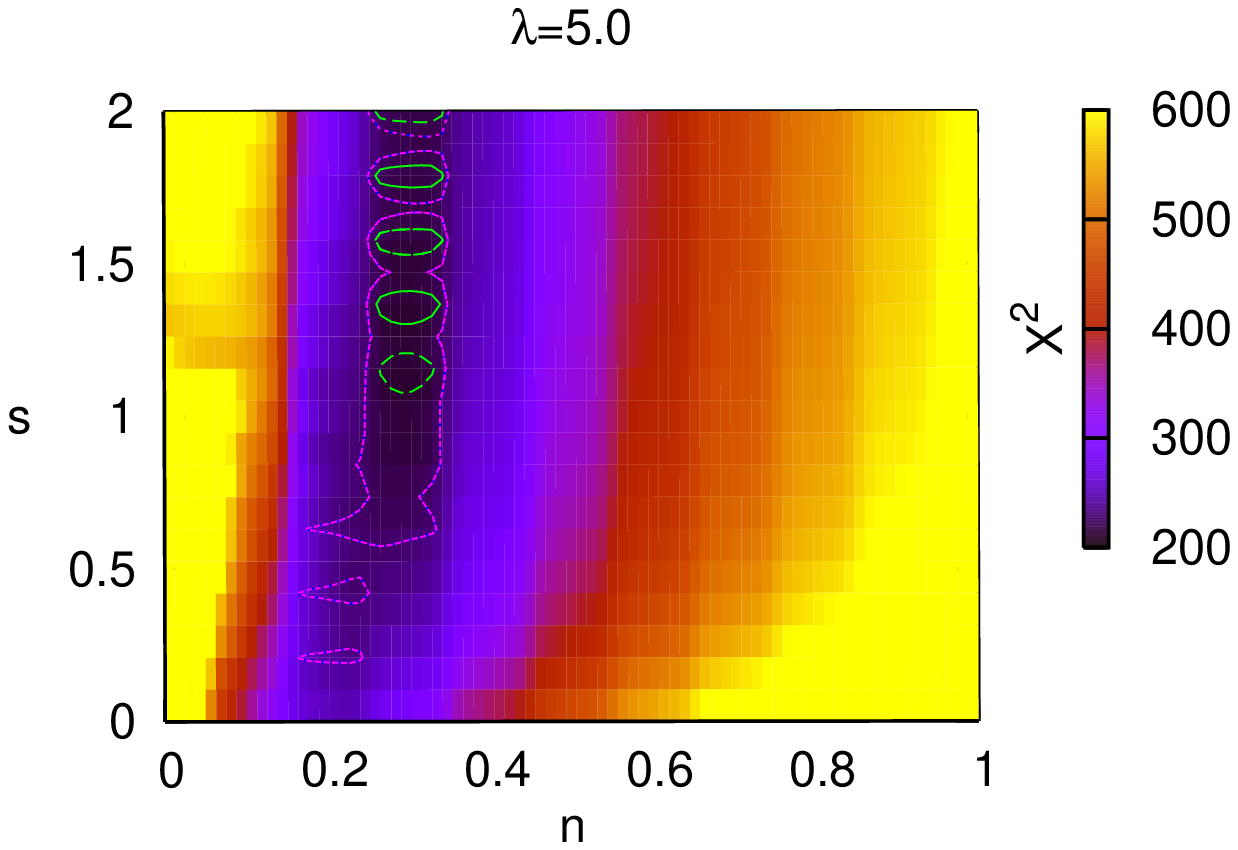} %
\includegraphics[width=5.5cm, angle=270]{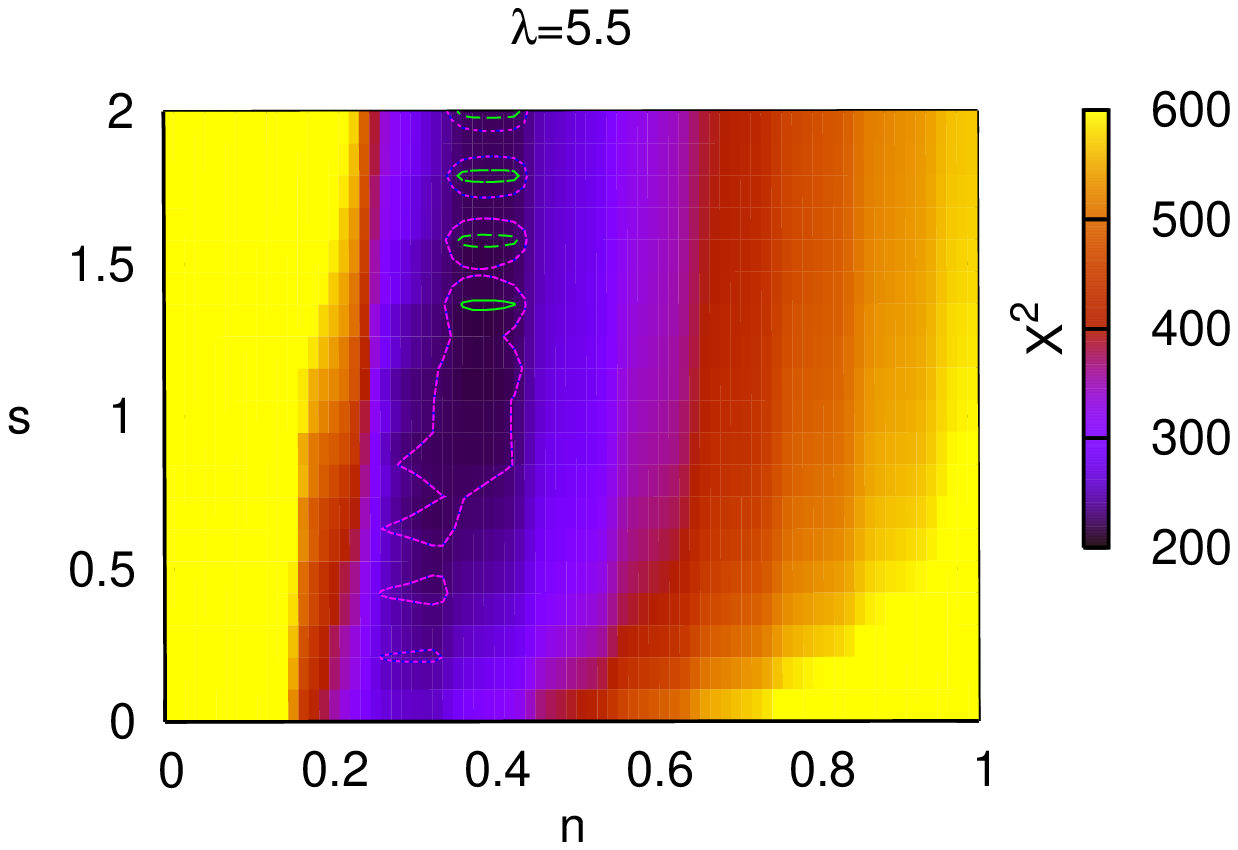}\hskip1cm%
\includegraphics[width=5.5cm, angle=270]{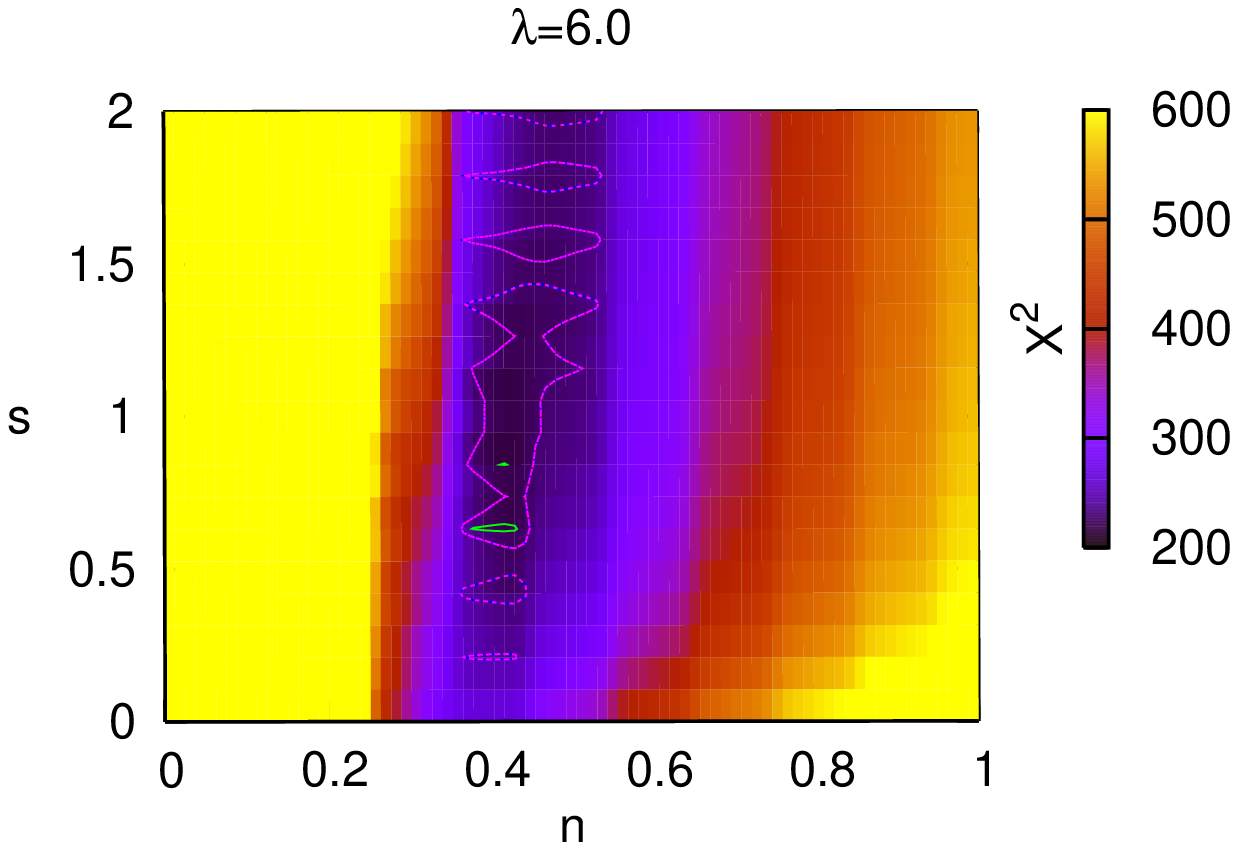} %
\includegraphics[width=5.5cm, angle=270]{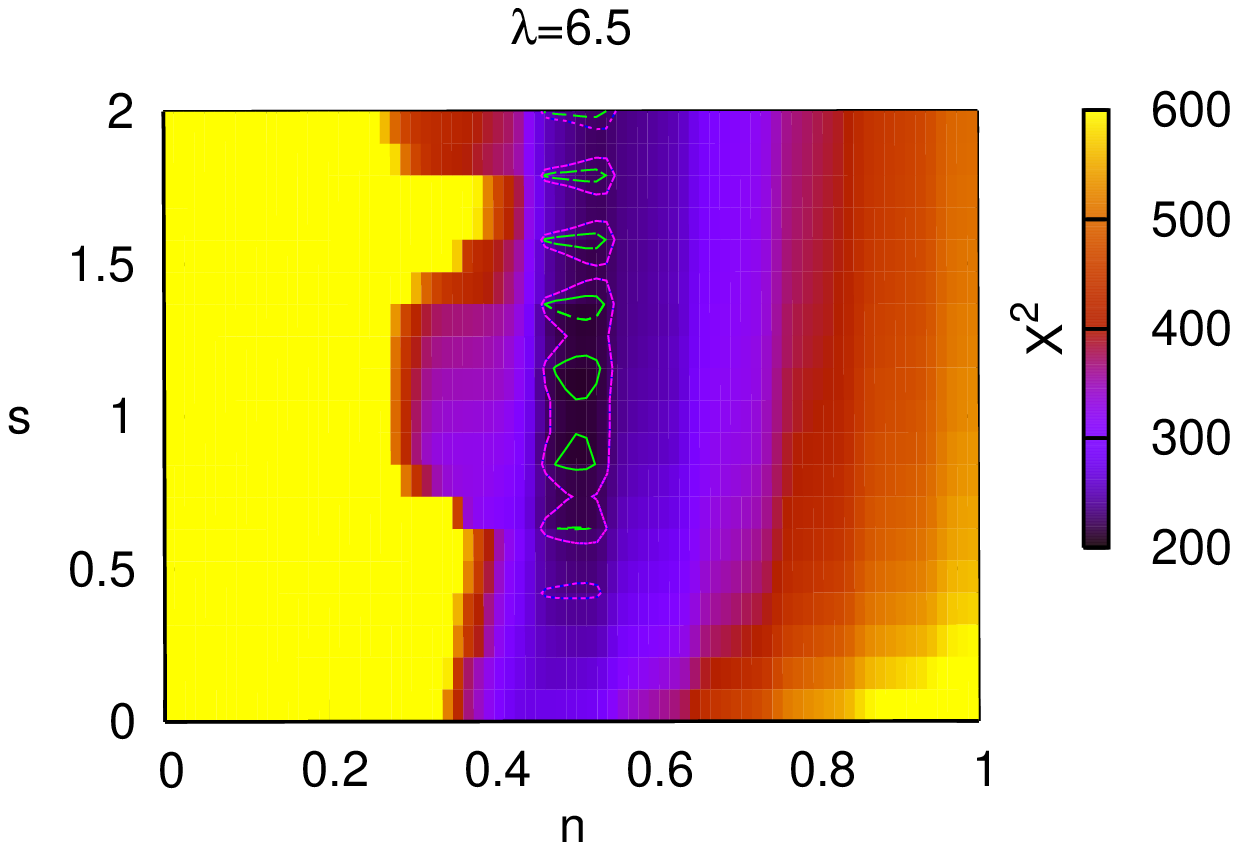}\hskip1cm%
\includegraphics[width=5.5cm, angle=270]{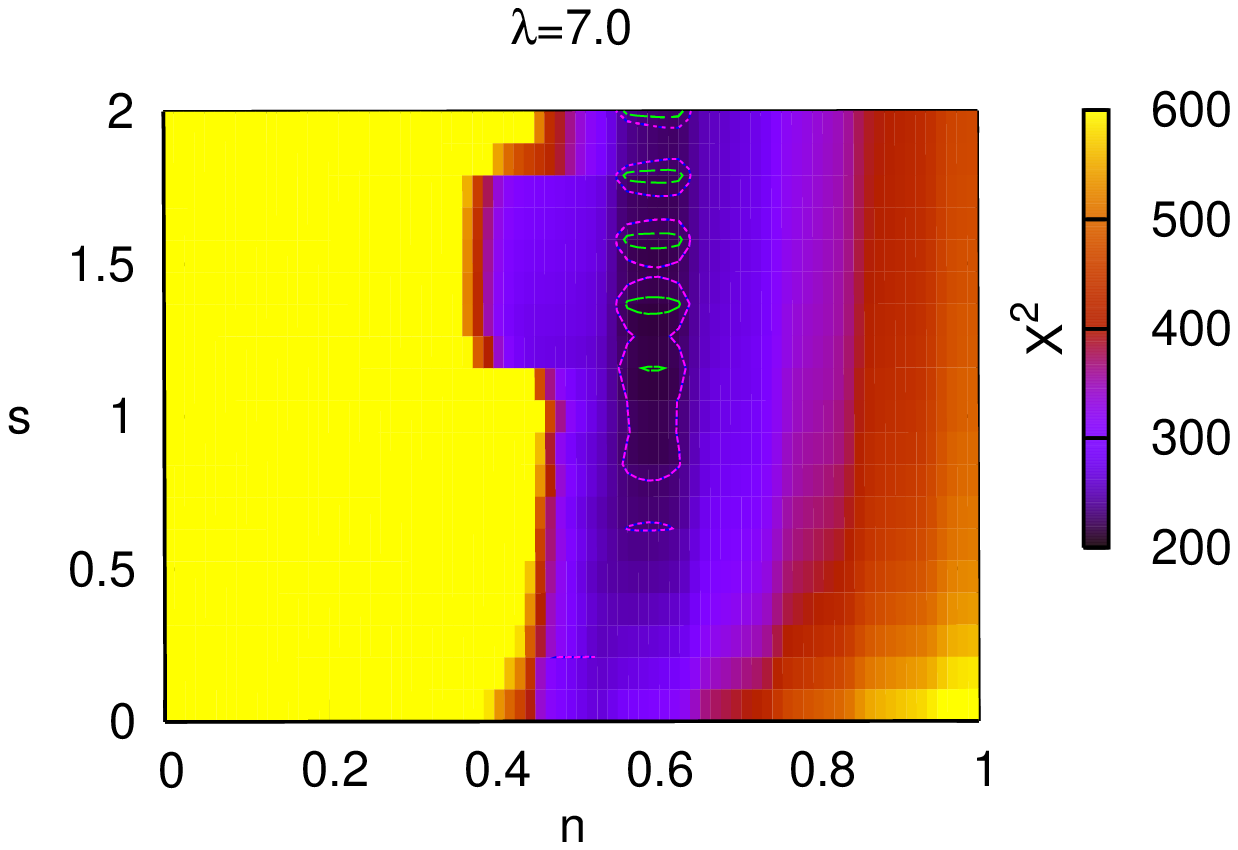} %
\includegraphics[width=5.5cm, angle=270]{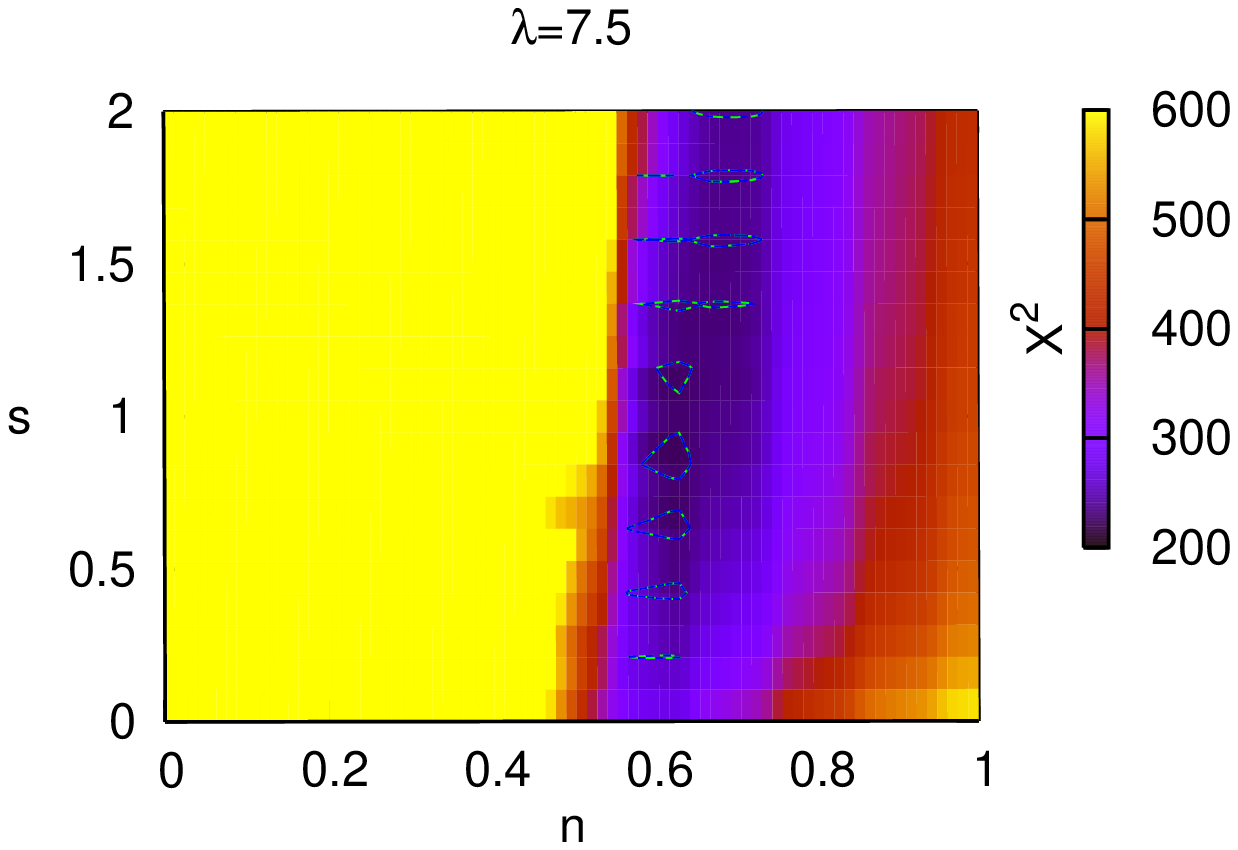}\hskip1cm%
\includegraphics[width=5.5cm, angle=270]{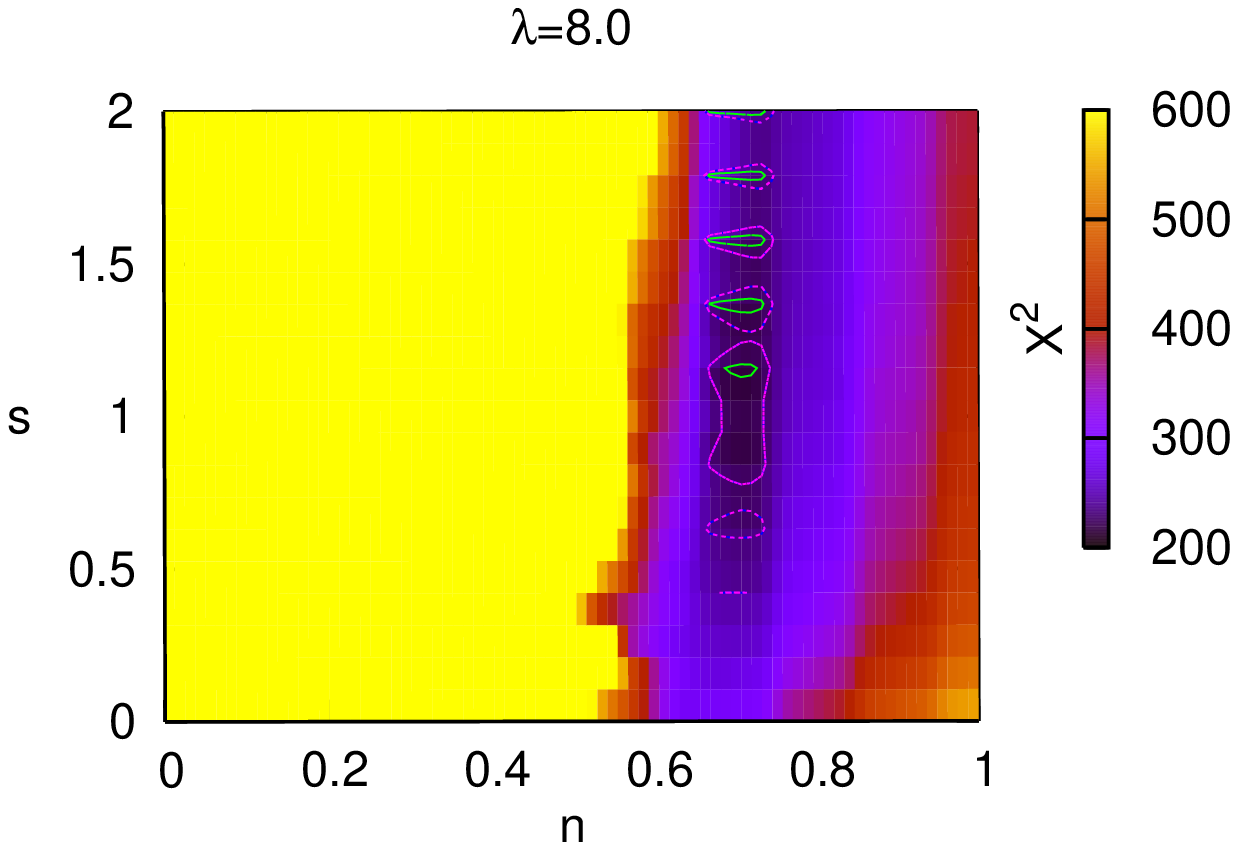}
\caption{The fit of the luminosity distance -- redshift relation
of the bulk viscous Chaplygin gas model to the Gold2006
\cite{gold06} supernova data. Different panels show models for
different $\lambda $ values, the inner (magenta) contours indicate
the 1-$\sigma $ confidence regions, the outer (turquoise) contours border the 2-$%
\sigma $ confidence regions. For all plotted values of $\lambda $
an accurate value of $n$ with a good fit can be established, while
the fit remains  acceptable for a wide range of the parameter
$s$.} \label{8panelmodel}
\end{figure}

Riess et al. \cite{gold06} have recently published a new set of
182 gold supernovae, including new Hubble Space Telescope (HST)
observations, and recalibrations of the previous measurements. We
applied the same tests to the Gold2006 data set \cite{gold06}, as
described in \cite{ger}, by taking $\gamma =0$. Here $n$ and $s$
were adjusted, and fixed values of $\lambda $=4.0, 4.5, 5.0, 5.5,
6.0, 6.5, 7.0, 7.5, 8.0 were taken into account. For any probed
values of $\lambda $, there exists a typical value of $n$, where a
wide range of $s$ offers a fit to the
supernovae data, as shown in Fig~\ref{8panelmodel}. As all panels in Fig.~\ref%
{8panelmodel} show, the acceptable fits occur in several disjunct
regions, which are slightly separated by $s$. The best-fit $n$
value was searched in the marginal projections of $n$ ($0<s<2$),
instead of finding a global minimum. The varying $s$ caused only
slight modifications of the fit, which
is characterized by the marginal projection of $s$ \textit{within} the 2-$%
\sigma $ confidence region.

The preferred $n$ values and the variation with $s$ for the probed
$\lambda$ values are represented in Table \ref{table1}. The first
column represent the values of $\lambda $ we have used for
comparison. The best fits for $n$ are represented in the second
column, while the best fit values for $s$ as well as the allowed
range of the parameter is represented in the third column. In the
cases $\lambda =5.5$ and $\lambda =8.0$, respectively,  a value of
$s>1$ is required, and the fit quality increases with increasing
$s$.


\begin{table}[h]
\begin{tabular}{|c|c|c|}
\hline
$\lambda $ & $n$ & $s$  \\
\hline
$4.0$ & $0.1$ & $0.8$ $(s>0.2)$ \\
\hline
$4.5$ & $0.2$ & $0.8$ $(s>0.3)$ \\
\hline
$5.0$ & $0.25$ & $0.8$ $(s>0.5)$ \\
\hline
$5.5$ & $0.35$ & $s>1$ \\
\hline
$6$ & $0.4$ & $1.7$ $(s>0.8)$ \\
\hline
$6.5$ & $0.5$ & $s>0.7$ \\
\hline
$7$ & $0.6$ & 1 $(s>0.7)$ \\
\hline
$7.5$ & $0.65$ & poor fit for any $s$ \\
\hline
$8$ & $0.7$ & $s>1$\\
\hline
\end{tabular}
\caption{The preferred $n$ values and the variation with $s$ for
the probed $\lambda$ values.}\label{table1}
\end{table}

These results show that the model behaves with some complexity in
the prediction of the luminosity distance, as the parameters do
not behave monotonically (with the exception of $n$). However, the
fitting ellipsoids are well-defined (see Fig. \ref{8panelmodel}),
and in general the predictions of the model fit well the supernova
data.

\section{Discussions and final remarks}

In the present paper we have considered the dynamics of a bulk
viscous Chaplygin gas filled flat homogeneous and isotropic
universe. We have derived and formulated the evolution equations
of the system, we have considered their behavior by using both
analytical and numerical techniques, and we have compared the
predictions of our model with the supernova data. The most
attractive feature of the Chaplygin gas is that it could explain
the main observational properties of the Universe without
appealing to an effective cosmological constant. Generally, the
obtained analytical and numerical solutions of the gravitational
field equations describes an accelerating universe, with the
effective negative pressure induced by the Chaplygin gas and the
bulk viscous pressure driving the acceleration.

From the equation of state of the Chaplygin gas with $\gamma =0$ it follows
that for the critical values $p_{c}$ and $\rho _{c}$ of the pressure and
density the parameter $w_{c}=p_{c}/\rho _{c}$ is given by $w_{c}=-B/\rho
_{c}^{n+1}-\Pi _{c}/\rho _{c}$. Evaluating this relation at the present time
when $\rho _{c}=\rho _{c0}$ gives $B=-w_{c0}\rho _{c0}^{n+1}-\Pi \left( \rho
_{c0}\right) \rho _{c0}^{n}$. The Chaplygin gas behaves like a cosmological
constant for $w_{c0}=-1$, which gives the relation between the constant $B$
and the present day value of the bulk viscous pressure as
\begin{equation}
B=\rho _{c0}^{n+1}\left[ 1-\frac{\Pi \left( \rho _{c0}\right) }{\rho _{c0}}%
\right] =\left( \frac{3H_{0}^{2}}{8\pi G}\right) ^{n+1}\left[ 1-\frac{\Pi
\left( \rho _{c0}\right) }{\rho _{c0}}\right] ,
\end{equation}%
where $H_{0}=3.24\times 10^{-18}h$ s$^{-1}$, $0.5\leq h\leq 1$, is the
Hubble constant \cite{PeRa03,Pa03}. Since $\Pi \left( \rho _{c0}\right) <0$,
the presence of the bulk viscous effects can significantly increase the
value of $B$.

By comparing the model with $\gamma =0$ to the Gold 2006 supernova
data, it turns out that a good agreement with these observations
can be established for a wide range of the power $\ s\in \left(
0.2,~2\right) $ which occurs in the phenomenological laws
(\ref{csi}), which characterize the bulk viscosity coefficient.
The other viscosity parameter $\alpha $ can be obtained from the
equation $3^{s-1}\alpha H_{0}^{2s-1}=1$, and
by choosing a value for the Hubble parameter. For $h=0.7$ ($%
H_{0}=2.268\times 10^{-18}$s$^{-1}$) we obtain $\alpha =\left(
6.2385\times 10^{-11}\text{s}^{-0.6},~2.8573\times
10^{52}\text{s}^{3}\right) $ for the above-established range of
the parameter $s$. As for the equation of state of the Chaplygin
gas, by taking into account the definition of $\lambda $, $\lambda
=B/3^{n}H_{0}^{2n+2}$, and for the same value of the Hubble
parameter, the confrontation with supernova data selects the pairs
$\left( n,~B\right) $ represented on Fig. \ref{nB}.

\begin{figure}[tbp]
\includegraphics[height=5cm]{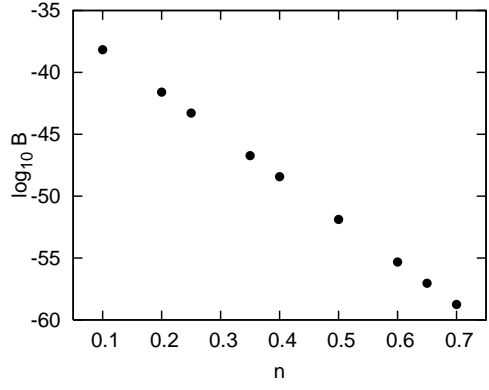}
\caption{The parameter values $B$ corresponding to the best fit values $%
\left( \protect\lambda ,~n\right) $ represented in a logarithmic
scale as a function of $n$.} \label{nB}
\end{figure}

Scalar fields are supposed to play a fundamental role in the
evolution of the early universe. The Chaplygin gas model can be
also described from a field theoretical point of view by
introducing a scalar field $\phi $ and a self interacting
potential $U(\phi )$, with the Lagrangian \cite{Ka01,Be02},
 \cite{Bi02}, \cite{Fa02}, \cite{De04}
\begin{equation}
L_{\phi }=\frac{1}{2}\dot{\phi}^{2}-U(\phi ).
\end{equation}

The energy density and the pressure associated to the scalar field $\phi $
associated to the bulk viscous Chaplygin gas are given by
\begin{equation}
\rho _{\phi }=\frac{\dot{\phi}^{2}}{2}+U\left( \phi \right) =\rho ,
\label{rho}
\end{equation}
and
\begin{equation}
p_{\phi }=\frac{\dot{\phi}^{2}}{2}-U\left( \phi \right) =\gamma \rho -\frac{B%
}{\rho ^{n}}+\Pi ,  \label{p}
\end{equation}
respectively.

The scalar field and the potential can be obtained from the equations
\begin{equation}
\phi \left( t\right) -\phi _{0}=\int_{t_{0}}^{t}\sqrt{\left( 1+\gamma
\right) \rho -\frac{B}{\rho ^{n}}+\Pi }dt,
\end{equation}
and
\begin{equation}
U\left( t\right) =\frac{1}{2}\left[ \left( 1-\gamma \right) \rho +\frac{B}{%
\rho ^{n}}-\Pi \right] ,
\end{equation}
respectively, where $\phi _{0}$ is an arbitrary constant of integration.

The dependence of the potential $U(\phi )$ on the scalar field $\phi $ is
represented in Fig.~\ref{FIG8}.

\begin{figure}[!ht]
\centering
\includegraphics{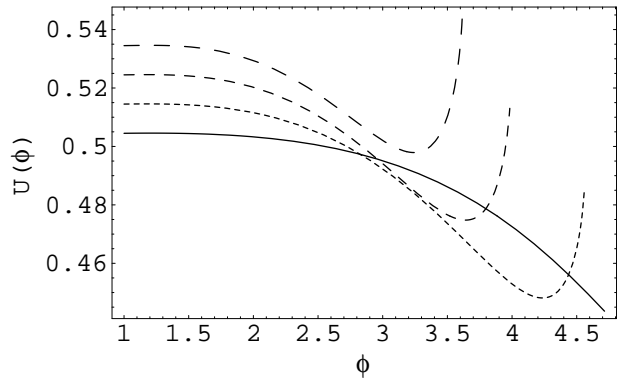}
\caption{The potential $U(\phi )$ of the viscous Chaplygin gas
associated scalar field as a function of the scalar field $\phi $
for a dust universe ($\gamma =0$), $n=0.1$, $s=1/4$, and for
different values of $\lambda _0$: $%
\lambda _0=0.01$ (solid curve), $\lambda _0=0.03$ (dotted curve), $%
\lambda _0=0.05$ (dashed curve) and $\lambda _0=0.07$ (long dashed
curve). } \label{FIG8}
\end{figure}

In conclusion, we have found that the viscous Chaplygin gas model offers a
real possibility for replacing the effective cosmological constant and to
explain the recent acceleration of the universe.

\acknowledgments  L\'{A}G was supported by the OTKA grant 69036
and by the Bolyai Grant of the Hungarian Academy of Sciences. KZ
was supported by the OTKA grant 69036. GMS was supported by the
Bolyai Grant of the Hungarian Academy of Sciences. TH was
supported by an RGC grant of the government of the Hong Kong SAR.


\begin{thebibliography}{99}
\bibitem{Pe99} S. Perlmutter et al., Astrophys. J. \textbf{517}, 565 (1998);
A. G. Riess et al., Astron. J. \textbf{116}, 109 (1998).

\bibitem{Ber00} P. de Bernardis et al., Nature \textbf{404}, 995 (2000); S.
Hanany et al., Astrophys. J. \textbf{545}, L5 (2000); D. N. Spergel et al.,
Astrophys. J. Supplement Series \textbf{170}, 377 (2007).

\bibitem{PeRa03} P. J. E. Peebles and B. Ratra, Rev. Mod. Phys. \textbf{75},
559 (2003).

\bibitem{Pa03} T. Padmanabhan, Phys. Repts. \textbf{380}, 235 (2003).

\bibitem{8} R. Caldwell, R. Dave and P. J. Steinhardt, Phys. Rev. Lett.
\textbf{80}, 1582 (1998).

\bibitem{quint} M. K. Mak and T. Harko, Int. J. Mod. Phys. D \textbf{11},
1389 (2002); C. R. Watson, R. J. Scherrer, Phys. Rev. D \textbf{68}, 123524
(2003); S. Matarrese, C. Baccigalupi, F. Perrotta, Phys. Rev. D \textbf{70},
061301 (2004); S. Nojiri and S. D. Odintsov, Phys. Rev. D \textbf{70},
103522 (2004); L. Perivolaropoulos, Phys. Rev. D \textbf{71}, 063503 (2005).

\bibitem{15} O. Bertolami and P. J. Martins, Phys. Rev. D \textbf{61},
064007 (1999).

\bibitem{16} A. A. Sen, S. Sen and S. Sethi, Phys. Rev. D \textbf{63},
107501 (2001).

\bibitem{17} S. Sen and A. A. Sen, Phys. Rev. D \textbf{63}, 124006 (2001).

\bibitem{IsSt76} W. Israel and J. M. Stewart, Phys. Lett. A\textbf{58}, 213
(1976).

\bibitem{11} L. P. Chimento, A. S. Jakubi and D. Pavon, Phys. Rev. D \textbf{%
62}, 063508 (2000).

\bibitem{MaHa03} M. K. Mak and T. Harko, Int. J. Mod. Phys. D \textbf{12},
925 (2003).

\bibitem{Paetal} T. Matos and A. Urena-Lopez, Class. Quantum Grav. \textbf{17%
}, L75 (2000); C. Wetterich, Phys. Rev. D. \textbf{65}, 123512; T.
Padmanabhan and T. R. Choudhury, Phys. Rev. D \textbf{66}, 081301 (2002).

\bibitem{Ka01} A. Kamenshchik, U. Moschella and V. Pasquier, Phys. Lett. B
\textbf{511}, 265 (2001).

\bibitem{Be02} M. C. Bento, O. Bertolami and A. Sen, Phys. Rev. D \textbf{66}%
, 043507 (2002).

\bibitem{Ka00} A. Kamenshchik, U. Moschella and V. Pasquier, Phys. Lett. B
\textbf{487}, 7 (2000).

\bibitem{Ka98} S. K. Kama, Phys. Lett. B \textbf{424}, 39 (1998).

\bibitem{Bi02} N. Bilic, G. B. Tupper and R. D. Viollier, Phys. Lett. B
\textbf{535}, 17 (2002).

\bibitem{No05} M. Novello, M. Makler, L. S. Werneck and C. A. Romero, Phys.
Rev. D \textbf{71}, 043515 (2005).

\bibitem{Chco} D. Carturan and F. Finelli, Phys. Rev. D \textbf{68}, 103501
(2003); R. Bean and O. Dore, Phys. Rev. D \textbf{68}, 023515 (2003); L. M.
G. Beca, P. P. Avelino, J. P. M. de Carvalho and C. J. A. P. Martins, Phys.
Rev. D \textbf{67}, 101301 (2003); M. C. Bento, O. Bertolami and A. A. Sen,
Phys. Rev. D \textbf{70} 083519, (2004); R. R. R. Reis, I. Waga, M. O.
Calvao and S. E. Joras, Phys. Rev. D \textbf{68} 061302 (2003); P. P.
Avelino, L. M. G. Beca, J. P. M. de Carvalho and C. J. A. P. Martins, JCAP
\textbf{0309} 002 (2003); G. M. Kremer, Phys. Rev. D \textbf{68} 123507
(2003); T. Multamaki, M. Manera and E. Gaztanaga, Phys. Rev. D \textbf{69},
023004 (2004); M. Szydlowski and W. Czaja, Phys. Rev. D \textbf{69} 023506
(2004); P. P. Avelino, L. M. G. Beca, J. P. M. de Carvalho, C. J. A. P.
Martins and E.J. Copeland, Phys. Rev. D \textbf{69} 041301 (2004).

\bibitem{Fa02} M. Makler, S. Quinet de Oliveira and I. Waga, Phys. Lett. B
\textbf{555}, 1 (2003); O. Bertolami, A. A. Sen, S. Sen and P. T. Silva,
Mon. Not. Roy. Astron. Soc. \textbf{353}, 329 (2004); J. V. Cunha, J. S.
Alcaniz, J. A. S. Lima, Phys. Rev. D \textbf{69}, 083501 (2004).

\bibitem{Be03} M. C. Bento, O. Bertolami and A. A. Sen, Phys. Lett. B
\textbf{575}, 172 (2003); M. C. Bento, O. Bertolami and A. A. Sen, Phys.
Rev. D \textbf{67}, 063003 (2003); L. Amendola, F. Finelli, C. Burigana, D.
Carturan, JCAP \textbf{0307}, 005 (2003).

\bibitem{De03} A. Dev, D. Jain and J. S. Alcaniz, Phys. Rev. D \textbf{67},
023515 (2003); A. Dev, D. Jain and J. S. Alcaniz, Astron. Astrophys. \textbf{%
417}, 847 (2004); P. T. Silva and O. Bertolami, Astrophys. J. \textbf{599},
829 (2003).

\bibitem{Al03} J. S. Alcaniz, D. Jain and A. Dev, Phys. Rev. D \textbf{67},
043514 (2003).

\bibitem{Cu04} J. V. Cunha, J. A. S. Lima and J. S. Alcaniz, Phys. Rev. D
\textbf{69}, 083501 (2004); Z.-H. Zhu, Astron. Astrophys. \textbf{423}, 421
(2004).

\bibitem{MaHa05} M. K. Mak and T. Harko, Phys. Rev. D \textbf{71}, 104022
(2005).

\bibitem{GiMe06} T. Giannantonio and A. Melchiorri, Class. Quant. Grav.
\textbf{23}, 4125 (2006).

\bibitem{Ber06} M. C. Bento, O. Bertolami, M. J. Reboucas and P. T. Silva,
Phys. Rev. D \textbf{73}, 043504 (2006).

\bibitem{BaGhKu07} R. Banerjee, S. Ghosh and S. Kulkarni, Phys. Rev. D
\textbf{75}, 025008 (2007).

\bibitem{GoWaWa07} Y. Gong, B. Wang and A. Wang, Phys. Rev. D \textbf{75},
123516 (2007).

\bibitem{WuYu07} P. Wu and H. Yu, Phys. Lett. B \textbf{644}, 16 (2007); P.
Wu and H. Yu, JCAP 03, 015 (2007).

\bibitem{HeSe07} M. Heydari-Fard and H. R. Sepangi, arXiv:0710.2666 (2007).

\bibitem{Ma} R. Maartens, \textsl{Causal thermodynamics in relativity},
astro-ph/9609119 (1996).

\bibitem{Ec40} C. Eckart, Phys. Rev. \textbf{58}, 919 (1940).

\bibitem{LaLi87} L. D. Landau and E. M. Lifshitz, \textit{Fluid Mechanics},
Butterworth Heinemann, Oxford (1987).

\bibitem{Is76}  W. Israel, Ann. Phys. {\bf 100}, 310  (1976).


\bibitem{HiLi89} W. A. Hiscock and L. Lindblom, Ann. Phys. \textbf{151}, 466
(1989).

\bibitem{HiSa91} W. A. Hiscock and J. Salmonson, Phys. Rev. \textbf{D43},
3249 (1991).

\bibitem{Ma95} R. Maartens, Class. Quantum Grav. \textbf{12}, 1455 (1995).

\bibitem{ChJa97} A. A. Coley and R. J. van den Hoogen, Class. Quantum Grav.
\textbf{12}, 1977 (1995); A. A. Coley and R. J. van den Hoogen, Phys. Rev. D
\textbf{54}, 1393 (1996); R. Maartens and J. Triginer, Phys. Rev. D \textbf{%
56}, 4640 (1997); L. P. Chimento and A. S. Jakubi, Class. Quantum Grav.
\textbf{14}, 1811 (1997); M. K. Mak and T. Harko, J. Math. Phys. \textbf{39}%
, 5458 (1998); M. K. Mak and T. Harko, Gen. Rel. Grav. \textbf{30}, 1171
(1998); M. K. Mak and T. Harko, Gen. Rel. Grav. \textbf{31}, 273 (1999); A.
Di Prisco, L. Herrera and J. Ibanez, Phys. Rev. D \textbf{63}, 023501
(2001); M. K. Mak and T. Harko, Europhys. Lett. \textbf{56}, 762 (2001);
Chiang-Mei Chen, T. Harko and M. K. Mak, Phys. Rev. D \textbf{64}, 124017
(2001); J. A. Belinchon, T. Harko and M. K. Mak, Class. Quant. Grav. \textbf{%
19}, 3003 (2002); T. Harko and M. K. Mak, Class. Quant. Grav. \textbf{20},
407 (2003).

\bibitem{De04} U. Debnath, A. Banerjee and S. Chakraborty, Class. Quant.
Grav. \textbf{21}, 5609 (2004).

\bibitem{law} G. L. Murphy, Phys. Rev. D \textbf{8}, 4231 (1973); V. A.
Belinskii and I. M. Khalatnikov, Sov. Phys. JETP \textbf{42}, 205 (1975); V.
A. Belinskii, E. S. Nikomarov and I. M. Khalatnikov, Sov. Phys. JETP \textbf{%
50}, 213 (1979).

\bibitem{ger} Z. Keresztes, L. \'{A}. Gergely, B. Nagy and G. M. Szab\'{o},
PMC Physics A \textbf{1}, 4 (2007); G. M. Szab\'{o}, L. \'{A}.
Gergely and Z. Keresztes, PMC Physics A \textbf{1}, 8 (2007); L.
\'{A}. Gergely, Z. Keresztes, G. M. Szab\'{o}, AIP Conference
Proceedings \textbf{957}, 391 (2007), \textit{arXiv:0709.0933}.

\bibitem{gold06} A. G. Riess, L.-G. Strolger, S. Casertano et al., \textit{%
New Hubble Space Telescope Discoveries of Type Ia Supernovae at }$z>1$%
\textit{: Narrowing Constraints on the Early Behavior of Dark
Energy}, to appear in Astrophys. J. \textbf{656} (2007);
\textit{astro-ph/0611572}.
\end{thebibliography}
\end{document}